\documentclass[cameraready]{Interspeech}

\title{Grounding Spoken LLMs in Multi-Speaker Audio \\ via Diarization Conditioning}

\author[affiliation={1,2}, orcid=0009-0000-4958-202X]{Alexander}{Polok}
\author[affiliation={2}, orcid=0000-0002-5358-1844]{Samuele}{Cornell}
\author[affiliation={1}, orcid=0000-0002-2225-5464]{Sathvik}{Udupa}
\author[affiliation={1}, orcid=0000-0002-8800-0210]{Jan}{Černocký}
\author[affiliation={2}, orcid=0000-0002-5970-8631]{\\Shinji}{Watanabe}
\author[affiliation={1}, orcid=0000-0002-4951-5908]{Lukáš}{Burget}

\address{
    $^1$ Speech@FIT, Brno University of Technology, Czechia \\
    $^2$ Language Technologies Institute, Carnegie Mellon University, USA
}

\email{ipoloka@fit.vut.cz}


\usepackage{comment}
\usepackage{makecell}
\usepackage{IEEEtrantools}
\usepackage{cite}

\begin{document}
\bstctlcite{IEEEexample:BSTcontrol}
\maketitle

\begin{abstract}
We propose diarization-conditioned spoken language models (SLMs), a strategy for extending SLMs to far-field multi-talker audio. Rather than adapting the decoder via Serialized Output Training, which risks catastrophic forgetting, we condition the acoustic encoder on diarization masks to extract target-speaker representations, keeping the decoder frozen. We instantiate this as Dixtral, integrating a Diarization Conditioned Whisper (DiCoW) encoder into the Voxtral SLM. On AMI, NOTSOFAR-1, LibriSpeechMix, and Mixer6, Dixtral outperforms Gemini 3.0 Flash, VibeVoice, and Voxtral Mini Transcribe V2 on speaker-attributed transcription by 29.0\%, 19.8\%, and 16.0\% absolute cpWER respectively. On a novel long-form multi-speaker QA benchmark, zero-shot Dixtral matches Gemini on far-field content understanding, and when fine-tuned surpasses both Gemini and Voxtral operating on close-talk across all tasks.

\end{abstract}

\section{Introduction}
\label{sec:intro}
Spoken Large Language Models (SLMs) are increasingly being more capable of multi-talker speech processing~\cite{peng2026vibevoiceasrtechnicalreport, huo2026tagspeechendtoendmultispeakerasr, mt_llm, yu2026moss} with some models now supporting long-form speaker attributed transcription. 
While not yet as performant as the modular pipelines~\cite{wang23_chime, niu24_chime, ye2023iacas, mitrofanov2024stcon, kamo24_chime, polok24_butjhu} that dominate 
current challenges such as CHiME-6-8~\cite{watanabe2020chime, CORNELL2026101901, zmolikova24_chime} and 
NOTSOFAR-1~\cite{abramovski2025summary, vinnikov24_interspeech}, 
they promise architectural simplicity and, crucially, zero-shot 
generalization to downstream tasks such as summarization and 
question-answering that modular systems cannot address.

The majority of these Large Language Models (LLM)-based systems rely on Serialized Output Training (SOT)~\cite{kanda20b_interspeech}, which concatenates multiple speakers' transcripts into a single token stream ordered by speech onset. While SOT and its variants~\cite{kanda22_interspeech, liang23e_interspeech, 10446059, kocour2025adaptingdiarizationconditionedwhisperendtoend} are effective for conventional encoder-decoder ASR, extending them to LLM decoders introduces a fundamental architectural challenge: SOT relies on special speaker-change tokens to manage overlapping speech. Accommodating these tokens requires expanding the LLM's vocabulary and retraining at least some parameters. This structural hurdle is further worsened by a distributional mismatch, as the interleaved, temporally-ordered sequences of SOT diverge significantly from the structured, single-speaker formats expected by instruction-tuned LLMs~\cite{shi2025serializedoutputpromptinglarge}.
Consequently, overcoming this mismatch requires decoder fine-tuning on large amounts of multi-speaker data. This in turn causes catastrophic forgetting of the model's reasoning, summarization, and question-answering capabilities~\cite{11151751, li-etal-2024-revisiting}---precisely the capabilities that motivate using an LLM rather than a task-specific end-to-end transcription model.

We argue that \emph{target-speaker (TS) extraction}~\cite{ts_extraction, wang20z_interspeech, medennikov20_interspeech} offers a more LLM-compatible formulation. Rather than forcing the decoder to manage multiple interleaved speakers, we condition the acoustic encoder to extract target-speaker representations. This ensures the LLM generates a single-speaker transcript, more closely matching the distribution it was pretrained on. Target-speaker extraction requires no special speaker-change or serialization tokens, leaving the LLM's vocabulary untouched, and, as we show, allowing the decoder to remain completely frozen. Furthermore, while executing inference separately for $S$ target speakers may intuitively seem computationally demanding due to repeated acoustic encoding, it actually provides superior scaling for spoken LLMs. Because the autoregressive generation step dominates computation and scales quadratically with sequence length, generating a joint multi-speaker transcript of length $SN$ results in a decoding complexity of $\mathcal{O}((SN)^2)$. In contrast, extracting each speaker independently requires $S$ separate decoding passes of length $N$, reducing the complexity to $\mathcal{O}(S \cdot N^2)$. As LLM backbones grow in size, this reduction in decoding overhead vastly outweighs the cost of running the smaller acoustic encoder multiple times. 


Current TS-ASR models generally rely on one of three primary paradigms: (1) integrating auxiliary speaker embeddings~\cite{Zili23_adapting, ma2024extending, moriya22_interspeech, speaker_beam}, (2) directly attending to target-speaker enrollment speech~\cite{meng24c_interspeech, guo2024sq, masumura23_interspeech}, or (3) leveraging diarization masks to guide the model's attention~\cite{meta_cat, wang25y_interspeech, polok2024targetspeakerasrwhisper}.
Building on the latter approach, we introduce Dixtral, a SLM for target-speaker transcription, summarization, and question-answering in multi-talker environments. Dixtral combines two complementary pretrained systems. The first is Voxtral~\cite{liu2025voxtral}, a SLM pairing a Whisper-based encoder~\cite{radford2023robust} with a Ministral~\cite{liu2026ministral3} decoder. The second is DiCoW~\cite{DiCoW}, a complete target-speaker ASR (TS-ASR) system featuring a diarization-conditioned Whisper encoder that isolates target-speaker acoustic representations prior to passing them to Whisper decoder.

We evaluate Dixtral on a benchmark comprising four multi-talker datasets: NOTSOFAR-1 (NSF1)~\cite{vinnikov24_interspeech}, AMI~\cite{Mccowan2005_ami}, LibriSpeechMix~\cite{kanda20b_interspeech}, and Mixer6~\cite{Mixer6}. Transcription performance across these datasets is measured using concatenated minimum-permutation word error rate, or cpWER~\cite{Neumann2023MeetEval}. Additionally, we evaluate the model on NSF-QA, a purposely made synthetic question-answering and summarization benchmark built on NSF1 featuring both content and paralinguistic questions.

Dixtral outperforms Gemini 3.0 Flash~\cite{gemini3flash2025, gemini}, 
VibeVoice~\cite{peng2026vibevoiceasrtechnicalreport}, and Voxtral Mini Transcribe V2~\cite{liu2026voxtralrealtime} on 
speaker-attributed transcription, achieving a macro-average cpWER 
of 15.4\% compared to 44.4\%, 35.2\%, and 31.4\% respectively. 
On a novel long-form multi-speaker QA and summarization benchmark, 
zero-shot Dixtral is able to match Gemini on far-field content understanding.  
When fine-tuned, Dixtral operating on far-field mixtures significantly surpasses 
both Gemini and Voxtral operating on close-talk single-speaker 
audio across all tasks, including paralinguistic queries on emotion 
and gender that remain inaccessible to cascaded ASR+LLM pipelines.

\begin{figure*}[t!]
    \centering
    \includegraphics[width=0.9\linewidth]{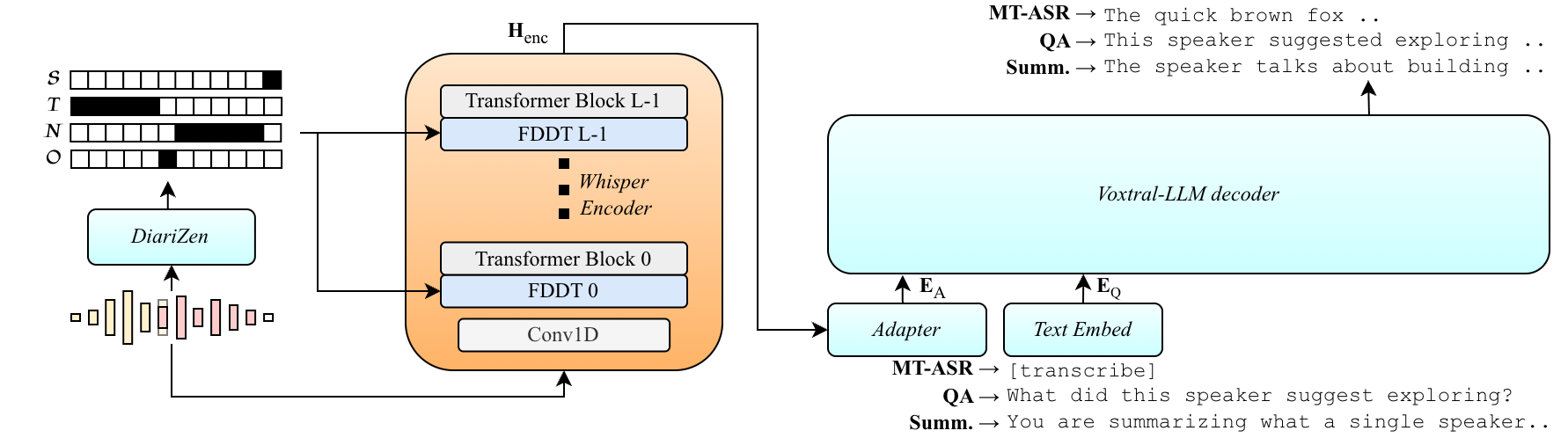 }
    \vspace{-2mm}
    \caption{Dixtral architecture with external DiariZen diarization conditioning. The DiCoW encoder conditions on STNO diarization masks via FDDT to extract target-speaker 
representations, which are projected into the frozen Mistral 
decoder. Only the encoder (orange) is trained; the diarization model, adapter and LLM (cyan) remain frozen.}
    \label{fig:architecture}
\end{figure*}

\section{Dixtral: Architecture}
\label{sec:method}

The overall architecture is illustrated in Figure~\ref{fig:architecture}. 
Dixtral builds upon Voxtral~\cite{liu2025voxtral}, a pretrained SLM 
comprising a Whisper~\cite{radford2023robust} encoder, a modality adapter, and 
a Ministral 3~\cite{liu2026ministral3} LLM-based decoder. 
While Voxtral demonstrates 
strong single-speaker speech understanding, it does not support 
speaker attribution. Our approach is to replace the encoder with 
a diarization-conditioned variant while keeping the LLM decoder 
frozen, a strategy applicable to any SLM. 
In this work we instantiate it with Voxtral and DiCoW~\cite{DiCoW} specifically 
because they share the same Whisper architecture, allowing us to 
initialize from DiCoW's pretrained weights and leverage Voxtral's 
existing cross-modal alignment.

\subsection{DiCoW: Diarization-Conditioned Whisper}
DiCoW is a complete end-to-end model trained specifically for TS-ASR. It performs target-speaker extraction by conditioning its acoustic encoder directly on diarization outputs. Because the DiCoW encoder is optimized to deliver target-speaker encodings to a standard Transformer decoder, these intermediate representations are inherently suitable for integration with the Voxtral pipeline.

The conditioning relies on frame-by-frame speaker activity probabilities, $d(s, t)$, where $s \in \{1, \dots, S\}$ indexes $S$ speakers and $t$ indexes time. For a given target speaker $s_k$, a Silence $\mathcal{S}$, Target $\mathcal{T}$, Non-target $\mathcal{N}$, and Overlap $\mathcal{O}$ mask, denoted as $\mathbf{STNO}_{s_k}$, is derived from the diarization output to capture four frame-level speech probabilities:
\begin{align}\label{eq:stno}
    p_{\mathcal{S}}^t  &= \prod_{s=1}^S (1 - d(s, t)), \quad
    p_{\mathcal{T}}^t  = d(s_k, t)  \prod_{\substack{s=1 \\ s \neq s_k}}^S (1 - d(s, t)) \nonumber \\
    p_{\mathcal{N}}^t  &= \left(1 - p_{\mathcal{S}}^t\right) - d(s_k, t), \quad
    p_{\mathcal{O}}^t  = d(s_k, t) - p_{\mathcal{T}}^t.
\end{align}

Instead of directly masking the input audio $\mathbf{X}$, these $\mathcal{STNO}$ probabilities are integrated through Frame-Level Diarization-Dependent Transformations (FDDT), that modulate the internal representations of each Transformer layer in the encoder. Each layer is augmented with four learnable diagonal affine transformation matrices, $(\mathbf{W}_i^l, \mathbf{b}_i^l)$, corresponding to the four $\mathcal{STNO}$ categories $i \in \{\mathcal{S}, \mathcal{T}, \mathcal{N}, \mathcal{O}\}$. 

The input to the encoder block at layer $l$ and frame $t$, denoted as $\mathbf{z}^l_t$, is transformed as a probabilistic blend weighted by the $\mathcal{STNO}$ probabilities:
\begin{equation}
\label{eq:fddt}
    \hat{\mathbf{z}}^l_t = \sum_{i \in \{\mathcal{S}, \mathcal{T}, \mathcal{N}, \mathcal{O}\}} (\mathbf{W}^l_i \mathbf{z}^l_t + \mathbf{b}^l_i)p^t_i.
\end{equation}

Crucially, the encoder does not simply ingest $\mathbf{STNO}_{s_k}$ at the input layer. Instead, the trainable parameters of the acoustic encoder, denoted as $\theta_{\text{enc}}$, interleave standard acoustic encoding with the FDDT mechanism. At each of the $L$ encoder layers, the intermediate representations $\mathbf{z}^l$ are adapted into $\hat{\mathbf{z}}^l$. After passing through all layers, the encoder yields the final high-level target-speaker acoustic representations $\mathbf{H}_{\text{enc}}$:
\begin{equation}
    \mathbf{H}_{\text{enc}} = \text{Encoder}_{\theta_{\text{enc}}}(\mathbf{X}, \mathbf{STNO}_{s_k}) \in \mathbb{R}^{T_{\text{enc}} \times d_{\text{enc}}},
\end{equation}
where $T_{\text{enc}}$ is the encoded sequence length and $d_{\text{enc}}$ is the encoder hidden dimension.

\subsection{Dixtral: Target-Speaker Reasoning Framework}
Dixtral integrates the layer-wise target-speaker extraction of the DiCoW encoder with the LLM reasoning capabilities of Voxtral. The full reasoning pipeline operates as follows:

\subsubsection{Conditioned Acoustic Processing}
The input audio and corresponding diarization mask are processed by the DiCoW encoder to produce $\mathbf{H}_{\text{enc}}$, as defined above. Crucially, this formulation allows for dynamic task switching. If the goal is to extract a specific speaker, the $\mathbf{STNO}_{s_k}$ probabilities are derived from diarization as defined in Equation~\eqref{eq:stno}. If the goal is to reason globally over the entire multi-talker audio (or if the input is a single-speaker audio), we simply set the target probability $p_{\mathcal{T}}^t = 1$ and all other probabilities to $0$ for all frames $t$, instructing the encoder to process the global acoustic context. Alternatively, supervision can depend on an external Voice Activity Detection (VAD) system.

\subsubsection{Modality Adapter}
To align the conditioned acoustic representations with the semantic space of the LLM, $\mathbf{H}_{\text{enc}}$ is passed through a modality adapter. This Multi-Layer Perceptron (MLP) adapter consists of two linear projections and a GELU nonlinearity to map the encoder embedding space to match the LLM embedding space: $E_{A}=\text{MLP}(\mathbf{H}_{\text{enc}}) \in \mathbb{R}^{T_{\text{enc}/4} \times d_{\text{llm}}}$, where $d_{\text{llm}}$ is the embedding dimension of the LLM.

\subsubsection{Prompting and Generation}
A discrete textual prompt $\mathbf{Q} = (q_1, q_2, \dots, q_K)$ of length $K$ (e.g., a specific question about the audio, a summarization request, or a standard transcription instruction) is embedded into $\mathbf{E}_Q \in \mathbb{R}^{K \times d_{\text{llm}}}$. The projected audio embeddings $E_A$ and the text prompt embeddings $E_Q$ are concatenated to form the final prefix sequence $\mathbf{U} = [E_A; E_Q]$. The LLM decoder then autoregressively generates the target text sequence $Y=(y_1, y_2, \dots, y_N)$ by predicting each subsequent token $y_t = \text{LLM}(\mathbf{U}, Y_{<i})$. Following this, the Cross-Entropy loss is computed only over the tokens $Y$ that follow the prompt: 
\begin{equation}
\mathcal{L}_{CE}=-\sum_{i=1}^{N}\log p(y_{i}|\mathbf{U}, Y_{<i}; \theta_{\text{enc}}).
\end{equation}

\section{Experimental Setup}
\label{sec:experimental_setup}


\subsection{Model Initialization}
To initialize Dixtral, we leverage the shared pretraining dynamics of Voxtral and DiCoW. Because Voxtral learned its projector with a frozen Whisper encoder~\cite{liu2025voxtral} and DiCoW learned its FDDT with a frozen decoder, combining them introduces minimal architectural friction. This compatibility allows for an initialization design where one can either replace the entire acoustic encoder with DiCoW's pretrained weights, or selectively inject only the pretrained FDDT parameters into Voxtral's encoder. 

Importantly, during fine-tuning, we \textbf{freeze both the LLM and the modality adapter}, updating only the acoustic encoder layers and the FDDT modules to learn the STNO conditioning. For our experiments, we build upon the Voxtral Mini 3B model (which utilizes a Ministral 3B backbone and Whisper large-v3 acoustic encoder).

\subsection{Tasks and Metrics}

\subsubsection{Multi-talker ASR}


We evaluate transcription performance using Concatenated 
Minimum-Permutation Word Error Rate 
(cpWER)~\cite{Neumann2023MeetEval} across four datasets: 
AMI~\cite{Mccowan2005_ami}, LibriSpeechMix 
(LSMix)~\cite{kanda20b_interspeech}, NOTSOFAR-1 
(NSF-1)~\cite{vinnikov24_interspeech}, and Mixer6 
(MX6)~\cite{Mixer6}. Among these, MX6 is not included in training and serves to assess out-of-domain generalization. We employ DiariZen~\cite{han25_interspeech} as the diarization backbone for both DiCoW and Dixtral; it serves as a reliable baseline, and the models' sensitivity to this specific system has already been studied~\cite{DiCoW}.


\begin{table}[t]
    \centering
    \caption{QA and summarization dataset statistics on NSF-1.}
    \label{tab:qa_stats}
    \resizebox{\columnwidth}{!}{%
    \begin{tabular}{lrrrrrr}
        \toprule
        & \multicolumn{4}{c}{Content QA} & Emotion QA & Summ. \\
        \cmidrule(lr){2-5}
        Split & Entity & Topic & Y/N & Detail & \\
        \midrule
        Train & 1,111 & 1,139 & 1,557 & 775 & 1,196 & 390 \\
        Dev   & 579   & 593   & 807   & 402 & 579   & 201 \\
        Eval  & 1,092 & 1,114 & 1,515 & 753 & 1,145 & 379 \\
        \bottomrule
    \end{tabular}%
    }
    \vspace{-0.5cm}
\end{table}

\subsubsection{Audio Question Answering (QA) and Summarization}

To evaluate SLMs' capabilities beyond transcription, we introduce a long-form audio, target-speaker QA and summarization benchmark on NSF-1 (Table~\ref{tab:qa_stats}).

The QA benchmark comprises two categories. \emph{Content QA} covers entity, topic, yes/no, and detail questions answerable from text alone (e.g., entity: ``Which city did this speaker mention as an example of good dog parks?'', topic: ``What main concern did this speaker raise about including water features in the park?'', yes/no: ``Did this speaker support adding a dog park to the design?'', and detail: ``What suggestion did this speaker make regarding elderly people?'').
\emph{Paralinguistic QA} includes emotion and speaker gender questions (e.g., “What was the perceived gender of this speaker?”) that require audio understanding. Emotion questions are generated by providing Gemini with the target speaker's transcript alongside utterance-level emotion labels (9 classes) from emotion2vec~\cite{ma-etal-2024-emotion2vec}, extracted from close-talk recordings with ground-truth segmentation, enabling questions grounded  in specific moments (e.g.\ ``Did this speaker sound enthusiastic when  discussing the new playground?''). Speaker gender labels are derived from  session metadata. Paralinguistic questions cannot be answered by cascaded systems that discard the audio signal, making them a test for end-to-end audio understanding. QA predictions  are scored by Gemini 2.5 Flash as a binary LLM judge (correct\,/\,incorrect) and we report accuracy. We acknowledge that using Gemini for both reference generation and judging introduces potential circularity; if anything, this may favor Gemini-generated outputs in Table~\ref{tab:qa_results}, making Dixtral's advantage over Gemini conservative.
No question references timestamps or segment boundaries, so each system must locate and reason over relevant information across the entire meeting.

For summarization, each system is prompted to produce a concise summary (${<}$50 words) of what the target speaker said during the meeting. Five reference summaries per speaker are generated from ground-truth transcripts using Gemini 2.5 Flash, and system outputs are evaluated with ROUGE-L~\cite{lin2004rouge}, taking the maximum across references to accommodate valid variation in phrasing. 
The full benchmark, generation scripts, and judge prompts are released as open source\footnote{\url{https://hf.co/datasets/popcornell/NSF-QA}}.

\subsection{Training Details}
To demonstrate viability for standard academic compute, we train on just eight 24GB A5000 GPUs. Following SE-DiCoW~\cite{polok2026sedicow}, models are trained for 20k steps with a peak learning rate of 6e-5 (5k warmup steps, cosine decay to zero). To process 2-minute utterances within these memory limits, we employ gradient checkpointing, bfloat16 precision, and 4 gradient accumulation steps to achieve an effective global batch size of 32. Two settings exceed this budget: for AMI decoding, we chunk each session into ~5-minute segments split at diarization-derived pauses; for QA and summarization fine-tuning, the long-form audio does not fit on the A5000s and is instead run on H100 GPUs. All training recipes, inference scripts, and code are open-sourced\footnote{\url{https://github.com/BUTSpeechFIT/Dixtral}}.

\section{Results}
\label{sec:results}

\begin{table}[t]
    \centering
    \caption{Concatenated Minimum-Permutation Word Error Rate (cpWER, \%) comparison across multi-talker datasets against specialized MT-ASR baselines and general purpose spoken LMs. DiCoW v3.3 and Dixtral results are obtained using diarization from DiariZen.}
    \label{tab:main}
    \setlength{\tabcolsep}{2pt}
    \begin{tabular}{l*{6}c|c}
        \toprule
        & \multicolumn{1}{c}{NSF-1} 
        & \multicolumn{1}{c}{AMI} 
        & \multicolumn{3}{c}{LSMix} & MX-6 & Avg. \\
        & Small & SDM & 1 & 2 & 3 & CH4 \\
        \midrule
        \multicolumn{7}{l}{\textit{Specialized MT-ASR Models}}\\
        \makecell[l]{Voxtral MTv2} & 54.4 & 42.3 & 2.0 & 28.2 & 42.3 & 19.4 & 31.4 \\
        VibeVoice & 35.8  & 33.7 & 2.1  & 50.8  & 72.8  & 16.0 & 35.2 \\
        \makecell[l]{DiCoW v3.3} & 26.6 & 18.6 & 1.8 & 3.1 & 21.7 & 11.9 & 14.0 \\
        \midrule
        \multicolumn{7}{l}{\textit{General Purpose Spoken LMs}}\\
        Gemini 3.0 Flash & 39.1 & 56.3 & 4.5 & 23.3 & 84.7 & 58.3 & 44.4 \\
        Dixtral & 29.1 & 19.8 & 2.1 & 3.6  & 23.5 & 14.4 & 15.4 \\
        \bottomrule        
    \end{tabular}
\end{table}

\begin{table}[t]
    \centering
\caption{Ablation study (cpWER $\%$, oracle diarization). \emph{Enc.\ swap}: full DiCoW encoder replaces Voxtral's. \emph{FDDT swap}: 
only FDDT parameters are injected. \emph{LoRA}: low-rank adaptation 
added to the decoder. \textit{QA+Summ ft}: model further finetuned for summarization and question answering.}
    \label{tab:ablations}
    \setlength{\tabcolsep}{2pt}
    \begin{tabular}{l*{6}c}
        \toprule
        & \multicolumn{1}{c}{NSF-1} 
        & \multicolumn{1}{c}{AMI} 
        & \multicolumn{3}{c}{LSMix} & MX6\\
        & Small & SDM & 1 & 2 & 3 & CH4  \\
        \midrule
        Dixtral wo/ swap & 26.3 & 19.8 & 1.9 & 2.3 & 3.6 & 12.6 \\
        
        w/ enc. swap & 26.4 & 21.0 & 1.9 & 2.4 & 4.6 & 20.4 \\
        w/ FDDT swap & 26.3 & 17.1 & 1.8 & 2.2 & 3.5 & 16.2 \\
        w/ LORA &  21.3 & 16.4 & 2.0 & 2.6 & 4.8 & 14.5 \\
        QA+Summ ft & 24.3 & 19.4 & 2.2& 3.1 & 5.6 & 26.1 \\
        \bottomrule        
    \end{tabular}
    \vspace{-0.5em}
\end{table}

\subsection{Target-Speaker Transcription Performance}
Table~\ref{tab:main} evaluates Dixtral against task-specific 
systems (DiCoW v3.3~\cite{polok2026sedicow}, 
VibeVoice~\cite{peng2026vibevoiceasrtechnicalreport}, Voxtral\,MTv2~\cite{liu2026voxtralrealtime}) and the general-purpose 
Gemini 3.0 Flash~\cite{gemini3flash2025,gemini}. On NOTSOFAR-1, 
Dixtral achieves 29.1\% cpWER, substantially outperforming 
VibeVoice (35.8\%)\footnote{We used the official open-sourced checkpoint and 
code from the Authors:\,\url{https://github.com/microsoft/VibeVoice}}, Gemini (39.1\%), and Voxtral MTv2 (54.4\%). On AMI, 
Dixtral reaches 19.8\%, narrowing the gap to the dedicated DiCoW 
acoustic model (18.6\%) while simultaneously supporting zero-shot 
downstream tasks. This performance extends out-of-domain to 
Mixer6 CH4 (14.4\% cpWER), outperforming all other LLM-based 
systems (Gemini, Voxtral\,MTv2 and Vibevoice). 

Table~\ref{tab:ablations} isolates architectural choices using oracle diarization. Under strict compute budgets, a full encoder swap (\emph{w/ enc.\ swap}) converges fastest (32.0\% NSF-1 cpWER at 2k steps, compared to 37.7\% for FDDT-only and 38.8\% for random initialization). However, over longer training runs, random initialization (\emph{wo/ swap}) or hot-swapping only FDDT parameters (\emph{w/ FDDT swap}) proves superior. Preserving Voxtral's original encoder weights prevents downstream embedding shifts, maintaining stable cross-modal alignment. Applying Low-Rank Adaptation (\emph{w/ LORA}) to the decoder yields the best overall ASR performance (21.3\% NSF-1, 16.4\% AMI) by absorbing subtle language characteristic shifts without passing destructively large gradients to the acoustic encoder. 
Finally, further fine-tuning for downstream reasoning 
(\emph{QA+Summ ft}) degrades pure transcription, most visibly 
on Mixer-6 (14.5\% $\to$ 26.1\%), as the encoder adapts to 
optimize for reasoning at the expense of verbatim accuracy. 
Joint ASR and QA+summarization training could mitigate this 
effect but is left for future work.

\subsection{Question Answering and Summarization}

Table~\ref{tab:qa_results} evaluates QA and summarization on 
NSF-QA. To isolate the contribution of diarization conditioning we compare Dixtral on far-field mixtures 
against its ``parent model'' Voxtral\,Mv1 on close-talk per-speaker audio (since it lacks any multi-speaker capability).
We additionally report Gemini 3.0 Flash as a reference. This latter, for far-field audio, is prompted with positional speaker descriptions (e.g. ``Consider the speaker who starts speaking third from the beginning'') as it lacks direct 
diarization conditioning.

Zero-shot after ASR-only training, Dixtral on far-field mixtures 
matches Voxtral on close-talk for Emotion QA (25.4\% vs.\ 25.5\%) 
and summarization (ROUGE-L 24.4 vs.\ 24.1), confirming that the 
frozen decoder's capabilities are preserved. Content QA is lower (54.6\% vs.\ 68.3\%) due to the harder far-field multi-speaker condition, though 
comparable to Gemini on far-field (55.1\%). 
Adding LoRA to the 
decoder for MT-ASR fine-tuning is a double-edged sword: it 
surfaces speaker-level attributes, pushing Gender QA to 73.3\%---on 
par with Gemini (74.1\%) and far above Voxtral's near-chance 
49.7\%---but degrades instruction following, occasionally producing verbatim transcriptions instead of summaries (15.4 ROUGE-L).

After explicit fine-tuning on NSF-QA, Dixtral on far-field audio 
surpasses both Voxtral and Gemini on close-talk across all 
categories, demonstrating the full potential of diarization conditioning when task specific data is also available. 


\begin{table}[t]
\centering
\caption{QA and summarization results on NSF-QA. Content QA is 
answerable from text; Emotion and Gender require audio 
understanding. Baselines are zero-shot; Dixtral (finetuned) is 
trained on NSF-QA to demonstrate the method's 
potential.}
\label{tab:qa_results}
\setlength{\tabcolsep}{1pt}
\footnotesize
\begin{tabular}{l ccc c}
\toprule
 & \multicolumn{3}{c}{\textbf{QA} (Accuracy $\uparrow$)} & \textbf{Summ.}\\
\cmidrule(lr){2-4}
\textbf{System} & Content & Emotion & Gender & ROUGE-L $\uparrow$ \\
\midrule
\multicolumn{5}{l}{\textit{Far-field}} \\
Dixtral (zero-shot)  & 54.6 & 25.4 & 43.2 & 24.4 \\ 
Dixtral (+LoRA, zero-shot) & 56.9 & 22.2 & 73.3  & 15.4 \\ 
Dixtral (finetuned)    & 73.0 & 47.6 & 95.5  & 41.4 \\
Gemini 3.0 Flash  & 55.1 & 29.3 & 74.1 & 23.7 \\
\midrule
\multicolumn{5}{l}{\textit{Close-talk}} \\
Voxtral\,Mv1       & 68.3 & 25.5  & 49.7 & 24.1 \\ 
Gemini 3.0 Flash    & 68.1  & 34.1 & 75.0 & 26.4 \\ 

\bottomrule
\end{tabular}
\vspace{-0.5em}
\end{table}


\section{Conclusion}
\label{sec:conclusion}

We proposed diarization-conditioned spoken LLMs, a general strategy 
for grounding spoken LLMs in multi-talker audio by conditioning the 
encoder while keeping the decoder frozen. Instantiated as Dixtral, 
this approach outperforms existing spoken LLMs on speaker-attributed 
transcription across four datasets. 
On QA and summarization, zero-shot Dixtral roughly matches Gemini on 
far-field for content QA (54.6\% vs.\ 55.1\%) and Voxtral on 
close-talk per-speaker oracle signal for summarization and emotion QA, showing that the frozen decoder retains its general-purpose capabilities. When 
fine-tuned, Dixtral surpasses both Voxtral and Gemini on close-talk 
across all tasks, including emotion and gender queries that cascaded pipelines 
cannot address. These findings suggest that diarization conditioning can be a promising 
path for extending spoken LLMs to far-field multi-speaker 
understanding. Future work includes end-to-end joint training with 
diarization, multilingual evaluation, and scaling in tasks, model 
size, and training data.

\section{Acknowledgements}
The work was supported by Ministry of Education, Youth and Sports of the Czech Republic (MoE) through the OP JAK project ``Linguistics, Artificial Intelligence and Language and Speech Technologies: from Research to Applications'' (ID:CZ.02.01.01/00/23\_020/0008518), and Brno Ph.D. Talent Scholarship Programme. Computing on IT4I supercomputer was supported by MoE through the e-INFRA CZ (ID:90254).

\section{Generative AI Use Disclosure}
Generative AI tools have been used to help revise and refine the manuscript.

\bibliographystyle{IEEEtran}
\bibliography{mybib}

\end{document}